\documentclass[10pt]{article}
\usepackage{ae}
\usepackage{aecompl}       
\usepackage[latin1]{inputenc}
\usepackage{geometry}
\usepackage{natbib}
\usepackage{graphics}
\usepackage{url}

\bibpunct[;]{(}{)}{;}{a}{}{;} 

\title{Asterias: a parallelized web-based suite for the analysis 
of expression and aCGH data}

\author{
Andreu Alibés$^1$, Edward R.~Morrissey$^1$, Andrés Cañada$^1$, Oscar M.~Rueda$^1$,\\
David Casado, Patricio Yankilevich$^1$, \\
Ramón Díaz-Uriarte$^{1, 2}$\\
}

\date{
\small{
$^1$Statistical Computing Team, Structural and Computational Biology Programme, Spanish National Cancer Center
  (CNIO), Melchor Fernández Almagro 3, \\
Madrid, 28029, Spain\\
}
}

\footnotetext[2]{Corresponding author. Email:
  \texttt{rdiaz02@gmail.com}. Phone: \texttt{+34-91-224-6900}}

\begin{document}
\maketitle

\begin{abstract}

  The analysis of expression and CGH arrays plays a central role in the
  study of complex diseases, especially cancer, including finding markers
  for early diagnosis and prognosis, choosing an optimal therapy, or
  increasing our understanding of cancer development and metastasis.
  Asterias (\url{http://www.asterias.info}) is an integrated collection of
  freely-accessible web tools for the analysis of gene expression and aCGH
  data. Most of the tools use parallel computing (via MPI) and run on a
  server with 60 CPUs for computation; compared to a desktop or
  server-based but not parallelized application, parallelization provides
  speed ups of factors up to 50.  Most of our applications allow the user
  to obtain additional information for user-selected genes (chromosomal
  location, PubMed ids, Gene Ontology terms, etc) by using clickable links
  in tables and/or figures. Our tools include: normalization of
  expression and aCGH data (DNMAD); converting between different types of
  gene/clone and protein identifiers (IDconverter/IDClight); filtering and
  imputation (preP); finding differentially expressed genes related to
  patient class and survival data (Pomelo II); searching for models of
  class prediction (Tnasas); using random forests to search for minimal
  models for class prediction or for large subsets of genes with
  predictive capacity (GeneSrF); searching for molecular signatures and
  predictive genes with survival data (SignS); detecting regions of
  genomic DNA gain or loss (ADaCGH).  The capability to send results
  between different applications, access to additional functional
  information, and parallelized computation make our suite unique and
  exploit features only available to web-based applications.

\end{abstract}

\vspace*{80pt}
\textbf{Running header:} Asterias: web-based gene expression
  and aCGH analysis.

\vspace*{20pt}

\newpage
\section{Introduction}

Gene expression data from DNA microarrays have had a central role in
the study of complex diseases, especially cancer and, in the last 10
years, hundreds of papers using gene expression data from microarray
studies of cancer patients have been published
\citep{Rhodes.Chinnaiyan}. However, the use of data from
microarray studies for early diagnosis and prognosis and to help to
choose an optimal therapy, or to increase our understanding of cancer
development and metastasis, faces several challenges \citep{
  Rhodes.Chinnaiyan, Michiels,
 Ransohoff2005}. Some of the most relevant challenges are the validation of the robustness and stability of analysis' results,
the biological interpretation of those results, and the
integration of information from other sources (e.g., functional
annotation). To approach these difficulties we need computationally
efficient applications that can analyze massive amounts of data, and
that make no compromises with the statistical rigor of the analysis.

A large number of web applications for genomic data are available but many
have been developed for a single task (e.g.,GEMS ---\citealt{WuKasif}---, for
biclustering; VAMPIRE ---\citealt{Hsiao2005}---, for differential gene
expression analysis, CAPweb ---\citealt{capweb}--- for aCGH analysis). From a user's point of
view, integrated suites can be much more appealing: they show a common,
unified interface, similar requirements regarding data formatting, and
allow the user to perform complete sets of analysis (e.g., starting from
data normalization, following with data merging by gene ID, and finishing
with the search of differentially expressed genes and class prediction
models).  Some suites include RACE \citep{PsarrosNAR},
MIDAW \citep{RomualdiNAR},  Gepas \citep{gepas3, gepas2, gepas1,
  gepas0}, and CARMAweb \citep{carma}. 

Our suite Asterias, as some of the other available suites, offers
integrated analysis, the possibility to use either the full suite or just
specific applications, and access to additional functional information.
Asterias' unique features are: a) Asterias is explicitly designed to take
advantage of web-based applications running on a multi-server site, by
using load-balancing and, more importantly, parallelized execution.
Parallel computing \citep{pachecoMPI, fosterMPI} can result in dramatic
decreases in the time a user must wait to obtain results (e.g., Pomelo II,
the parallelized version of Pomelo ---\citep{gepas2, gepas1}---, can achieve
speed ups of factors up to 50 in our computing cluster). By itself, this
efficient use of a multi-server infrastructure makes Asterias unique. b)
Asterias leads to a careful examination of the problem of multiple solutions.
Studies about class prediction (e.g., cancer vs. non-cancer) with genomic
data have shown repeatedly \citep{Michiels,yo-azuaje, yo-rf, EinDor,
  Somorjai2003, pan-pnas, Yeung.Bumgarner2003} that many problems have
multiple "solutions" (sets of genes and models) with equivalent predictive
capacity. All the prediction tools in our suite Asterias (Tnasas, SignS,
GeneSrF) provide detailed reports about multiple solutions, by using
either cross-validation or the bootstrap. c) We emphasize careful
testing of our applications and, a unique feature in web-based
applications, provide complete source code of our automated testing
procedures. d) Asterias integrates tools that cover the whole range of needs of
many labs and researchers (normalization, filtering and missing value
imputation, differential gene expression, class prediction, survival
analysis, and aCGH analysis), being the only suite that incorporates
searching for large sets of predictive genes (GeneSrF) and prediction of
survival data (SignS).

\vspace*{20pt}

\begin{figure}[h!]
\begin{center}

{\resizebox{\columnwidth}{!}{%
\includegraphics{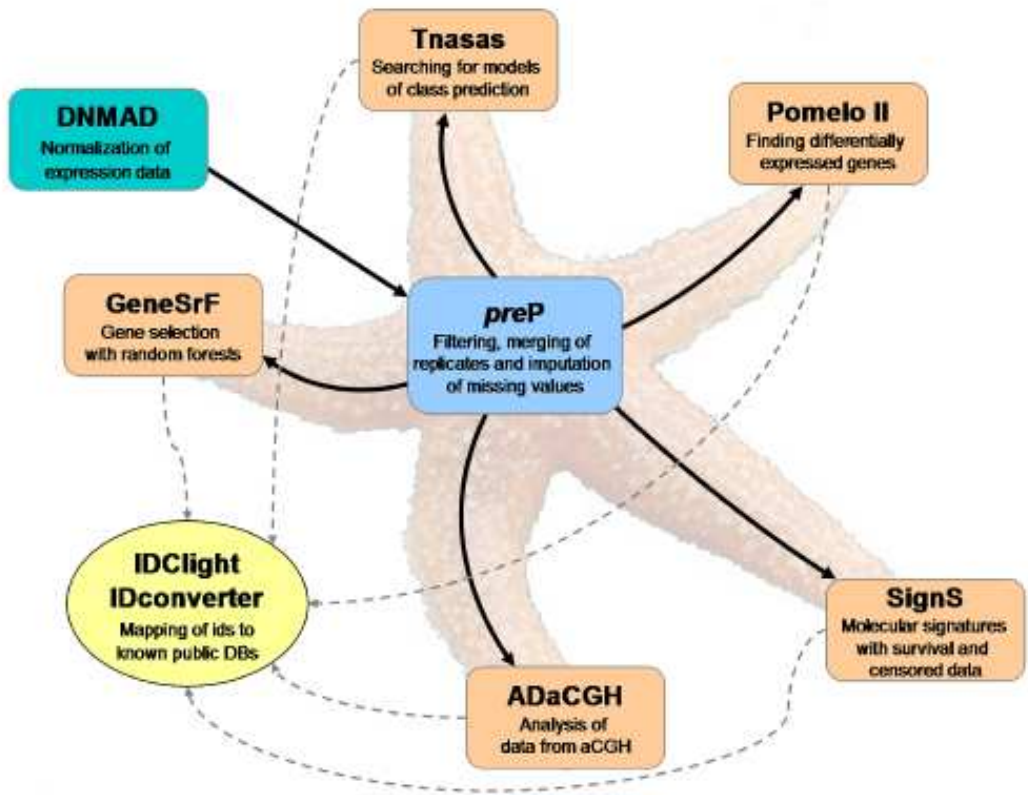}}}

\caption{\label{asterias.fig} Asterias suite. Relationships between the
  applications currently available in Asterias. An arrow indicates the
  possibility of automatically transferring the output from one
  application (origin of the arrow) as input for another application (end
  of row). All applications can also be accessed independently. Photo
  credit: the starfish is a modified image taken from the Wikipedia entry
  for Asterias (\texttt{http://en.wikipedia.org/wiki/Image:Asterias\_rubens.jpg}), and belongs to Hans Hillewaert.}
\end{center}
\end{figure}

\section{Functionality}

Asterias provides an integrated set of freely-available tools that
allows for comprehensive analysis of expression and aCGH data, from
normalization to searching for class and survival prediction models
and integration of additional functional information.  Figure
\ref{asterias.fig} shows the applications and their relationships. All
tools are accessible from preP, but can also be accessed
directly, and preP can be accessed either directly or from DNMAD. The
functionality and analysis provided by each application are:

\begin{description}
\item[DNMAD] Diagnosis and normalization of array data (both expression and
  aCGH). 
  \begin{itemize}
  \item Diagnostic plots to identify possible spatial patterns, arraying
    problems, and differences in spread among arrays and subarrays.
  \item Print-tip and global loess-based normalization.
  \item Use of flags to determine points to exclude and points to
    normalize but not use for determining the normalization curve.
  \item Three options for background correction.
  \item User-specified color ratio (red(Cy5)/green(Cy3) vs. green(Cy3)/red(Cy5)).
  \item Input as GPR files or custom formatted files, and upload of
    uncompressed or compressed files.
  \end{itemize}
  
\item[preP] Preprocessing of array data. 
  \begin{itemize}
  \item Filtering genes with missing data.
  \item Data imputation using KNN \citep{troya}.
  \item Merging of replicate spots in the array.
  \item Elimination of constant genes/clones/spots.
  \end{itemize}

\item[IDconverter] Mapping of clone, gene and protein ids to known public 
  databases.
    \begin{itemize}
  \item Output ids: 8 gene ids, 3 clone ids, 5 protein ids, plus PubMed
    references, GO terms, and KEGG and Reactome pathways. 
    Chromosomal location from two sources.
  \item Several output formats: HTML, tab separated text file and
    spreadsheet file.
  \end{itemize}
\item[IDClight] Same as IDconverter, but input coming directly from URL.

\item[Pomelo II] Finding differentially expressed genes.
  \begin{itemize}
  \item For differential expression associated to class differences (using
    t-test, paired t-test,  or ANOVA), a continuous variable (linear regression), or
    survival time (via Cox models).
  \item Unadjusted and FDR-adjusted p-values.
  \item For t-test, ANOVA, and regression, p-values can be obtained by data
    permutation.
  \item Empirical bayes moderated statistics for t-test and ANOVA.
  \item Addition of clinical covariates in linear models.
  \item Heatmaps with gene dendrograms of user-selected subsets of genes
    (filtering by statistic, absolute value of statistic, p-value and
    adjusted p-value and number of genes).
  \end{itemize}

\item[Tnasas] Searching for models of class prediction.
  \begin{itemize}
  \item Five different class-prediction algorithms (support vector
    machines, nearest neighbor, discriminant analysis, random forest, and
    shrunken centroids).
  \item Three different gene ranking methods (between-to-within sums of
    squares ---F-ratio---, Wilcoxon statistic, random forest).
  \item Honest assessment of prediction error rate using double
    cross-validation.
  \item Assessment of the relationship between number of genes in class
    prediction models and error rate.
  \item Comprehensive analysis of stability of solutions for both ``best''
    number of genes and identity of selected genes.
  \end{itemize}

\item[GeneSrF] Gene selection for classification problems using random
  forest. Targeted towards identifying both small, non-redundant sets of
  genes with good predictive performance (as explained in \citealt{yo-rf})
  as well as large sets of genes (including redundant genes) related to
  the outcome of interest.

  \begin{itemize}
  \item Honest assessment of prediction error rate using the bootstrap.
  \item Assessment of the relationship between number of genes in class
    prediction models and error rate.
  \item Comprehensive analysis of stability of solutions for both ``best''
    number of genes and identity of selected genes and selection
    probability plots.
  \item Importance spectrum plots and variable importances, to determine
    the relevance of genes.
  \end{itemize}

\item[SignS] Molecular signatures and gene selection with survival and censored data.
  \begin{itemize}
  \item Implements a method that uses a combination of gene filtering,
    clustering and survival model building (FCMS), very similar to the one used
    in \citet{DaveFCMS}.

  \item Honest assessment of model quality using cross-validation.
  \item Full details on models fitted and steps used, including
    detailed dendrograms, steps of
    variable selection, and correlation between signatures for FCMS.
  \item Comprehensive analysis of stability of solutions for both ``best''
    number of genes, identity of selected genes, and signatures.
  \item Option to use validation data to obtain assessments of model quality.
  \end{itemize}

\item[ADaCGH] Analysis of data from aCGH: calling gains and losses and
  estimating the number of copy changes in genomic DNA.
  \begin{itemize}
  \item Implements four methods that have been shown to perform well in
    previous studies \citep{Will.Frid2005, Lai.Park2005, PSW}: circular binary segmentation
    \citep{CBS}, wavelet-based smoothing  \citep{WaveletCGH}, Price-Smith-Waterman
    SW-ARRAY \citep{PSW}, and analysis of copy errors (the same method
    as implemented in CGH Explorer \citep{ACE}).
  \item Diagnostic plots and overimposed plots to help determine
    suitability of methods and number of levels of gain/loss.
  \end{itemize}

\end{description}

\section{Implementation}

\subsection{Software and hardware infrastructure}

Asterias runs on a computing and web-serving cluster with 30 nodes,
each with two Xeon CPUs. This cluster uses Debian GNU/Linux as OS. We use Apache as web
server, with web service load-balanced using Linux Virtual Server
(LVS); because most computations are parallelized using MPI (see
below), we use round-robin for web-service load-balancing.
High-availability is achieved using redundancy in both LVS (two
directors monitored with heartbeat) and storage (via a set of custom
scripts). The database server for IDconverter and IDClight is MySQL.

All applications (except preP, DNMAD and Tnasas) are parallelized
using the LAM/MPI implementation (\url{http://www.lam-mpi.org}) of
MPI. Pomelo II is parallelized in C++, whereas the rest of the
applications are parallelized in R using the library
Rmpi (\url{http://www.stats.uwo.ca/faculty/yu/Rmpi}), and snow
(\url{http://www.stat.uiowa.edu/~luke/R/cluster/cluster.html}) or
papply (\url{http://cran.r-project.org/src/contrib/Descriptions/papply.html}). The
MPI universe is created new for each run of each application, and the
actual nodes to use in the MPI universe are determined at run-time
after excluding possible non-responding nodes. This ensures that MPI
can be used even if a node fails or is taken down for
maintenance. When the parallelization does not involve all CPUs in the
cluster, the CPUs used in the MPI universe are balanced: the
configuration file for MPI depends on the master node of a run (and
the master node is the one where the Apache process runs, which is
balanced  by LVS).

\subsection{Software}

Computations are carried out using R and C/C++, either alone o called from
R. CGIs, data entry verification, MPI and cluster monitorization, and
application counter and monitorization are written with Python, except for
DNMAD (which use Perl). JavaScript is used for some of the dynamic output
(collapsible trees, some clickable figures, and Ajax). Further details
about design, implementation, and software and hardware organization, of
interest mainly to developers, are provided at the Asterias project's page
(\url{http://bioinformatics.org/asterias/wiki/Main/DevelopersDocumentation}).
As many other popular tools, we make extensive use of R and BioConductor
packages, but many functions have been rewritten to allow for parallel
computing. Full details on R/BioConductor packages used are provided on
the help pages for each application.

\subsection{Testing}
Testing that applications work as expected (see, e.g., \citealt{oxymoron})
is an integral part of the
development of Asterias. For most applications, a suite of tests, which
use the FunkLoad tool (\url{http://funkload.nuxeo.org/}), is available
from the the Asterias download page
(\url{http://bioinformatics.org/asterias/wiki/Main/DownloadPage}) or from
Launchpad (\url{https://launchpad.net/projects/asterias}). By using these
tests we verify CGIs (including JavaScript), numerical output, the
handling of error conditions and incorrectly formatted input files, and
the setting up of MPI universes. For Pomelo II (currently the application
that uses the most JavaScript and Ajax) we have also built tests
(available from \url{http://pomelo2.bioinfo.cnio.es/tests.html}), using
Selenium (\url{http://www.openqa.org/selenium/}), that verify that the
application runs correctly under different operating systems and browsers.

\subsection{Users, maturity and bug-tracking}
Asterias is a mature application suite, with a large number of users. Some
of the applications that form part of Asterias have been running for
almost three years (e.g., DNMAD, launched on October 2003) and the most
recent applications (IDClight, preP) have been running since January 2006.
The number of average daily uses (in the six-month period from
1-March-2006 to 1-September-2006) ranges from 5 per day for Pomelo II to
0.5 per day for SignS (Tnasas: 0.8; ADaCGH: 1.85; GeneSrF: 1; DNMAD:
3.88). For IDconverter the average daily uses are about 75 (IDClight uses
are over 500, but each counted use involves a single identifier).
Please note that the above are successful uses (i.e., only runs with
validly formatted data sets are counted). Asterias now includes a
bug-tracking and feature-requests page at
\url{http://bioinformatics.org/bugs/?group_id=630}.

\section{User interface}

\begin{table}[h!]

\caption{\label{input.output.table} Summary input and output for each
  application from the Asterias suite.}
\vspace*{12pt}


{ \footnotesize\small\footnotesize
\begin{tabular}{p{1.4cm}p{4cm}p{4cm}p{4cm}}
\hline
Application & Input & \multicolumn{2}{c}{Output}\\
&                                        &Tables and data sets&Figures\\
\hline
DNMAD      &GPR or custom format                           &Normalized log-ratios; A-values                           & Diagnostic plots  \\
\\
preP       &DNMAD output or EDF$^1$            &Post-processed EDF, summary statistics                    & \\
\\
Pomelo II  &preP output or EDF, class indicator, 
            survival time and status                       &Differential expression statistics and p-values           &Heatmap with gene dendrogram\\
\\
Tnasas     &preP output or EDF, class indicator            &Error rates, selected genes, stability assessments        &Cross-validated error rates vs. number of genes\\
\\
GeneSrF    &preP output or EDF, class indicator            &OOB predictions, error rates, selected genes, stability 
                                                                              assessments                             &OOB error vs. number of genes, OOB predictions, importance spectrum, 
                                                                                                                                selection probability plots\\
\\
SignS      &preP output or EDF, survival time and status; 
                       optional validation files           &Single-gene statistics and p-values, 
                                                             CV predictions, model results and parameters, 
                                                               stability assessments                                 & Survival plots, dendrograms, partial-likelihood plots.\\
\\
ADaCGH     &preP output or EDF and chromosomal location 
            (e.g., from IDconverter)                      &Genes and segmented regions, summary statistics           & Diagnostic plots, chromosome and genome segmented plots \\
\\
IDconverter&identifiers                                   &Mapped
identifiers (gene, clone, protein), chromosomal location, PubMed
                                                              abstracts,
                                                              GO terms, pathways.                             &\\                                                           
\\
IDClight   &URL$^2$                           &Same as IDconverter                                       &\\

\hline
\end{tabular}

$^1$ EDF: expression (or genomic) data file. See text for details.\\
$^2$ For example, \verb*$http://idclight.bioinfo.cnio.es/idclight.prog?idtype=ug&id=Hs.100890&org=Hs$
}

\end{table}

\subsection{Input}

All applications use plain text files, with tab-separated columns for
input. Missing values, in the applications that accept them, can be
specified as either ``NA'' or by not filling the corresponding entry. The
expression data files (\textbf{EDF}) (such as those returned by preP 
and used by all applications, except DNMAD and IDconverter)
are formatted with genes in rows and patients or arrays in columns. The
first column should be a column of identifiers, which can be of arbitrary
length and include any character except tab (since tab is used for column
separation).  The array data can include a row with array/subject
identifiers. It can also include an arbitrary number of comment lines
(all lines with a ``\#'' in the first column) anywhere in the
file. Comment lines are a convenient way to record all transformations
suffered by a file in DNMAD and preP.

DNMAD, IDconverter and IDClight have unique data entry
requirements/flexibilities. For IDconverter the entry is a column of one
or more identifiers. IDClight is designed for automated, programmable
access, and accepts ``data entries'' via URL (see example in Table
\ref{input.output.table}). DNMAD accepts either files in GPR format
(\url{http://www.moleculardevices.com/pages/software/gn_genepix_file_formats.html#gpr})
as produced directly by many microarray scanners, or non-GPR files, if
they have a specified set of columns (see DNMAD help). 

The analysis applications need additional files: class information
(e.g., Pomelo II), survival time and censored status (SignS, Pomelo II),
and chromosome location information (ADaCGH) such as is returned by
IDconverter. All these input files are also tab-separated, plain-text
files. Further details are shown in Table \ref{input.output.table}.

\subsection{Output}

\begin{figure}[h!]
\begin{center}

{\resizebox{\columnwidth}{!}{%
\includegraphics{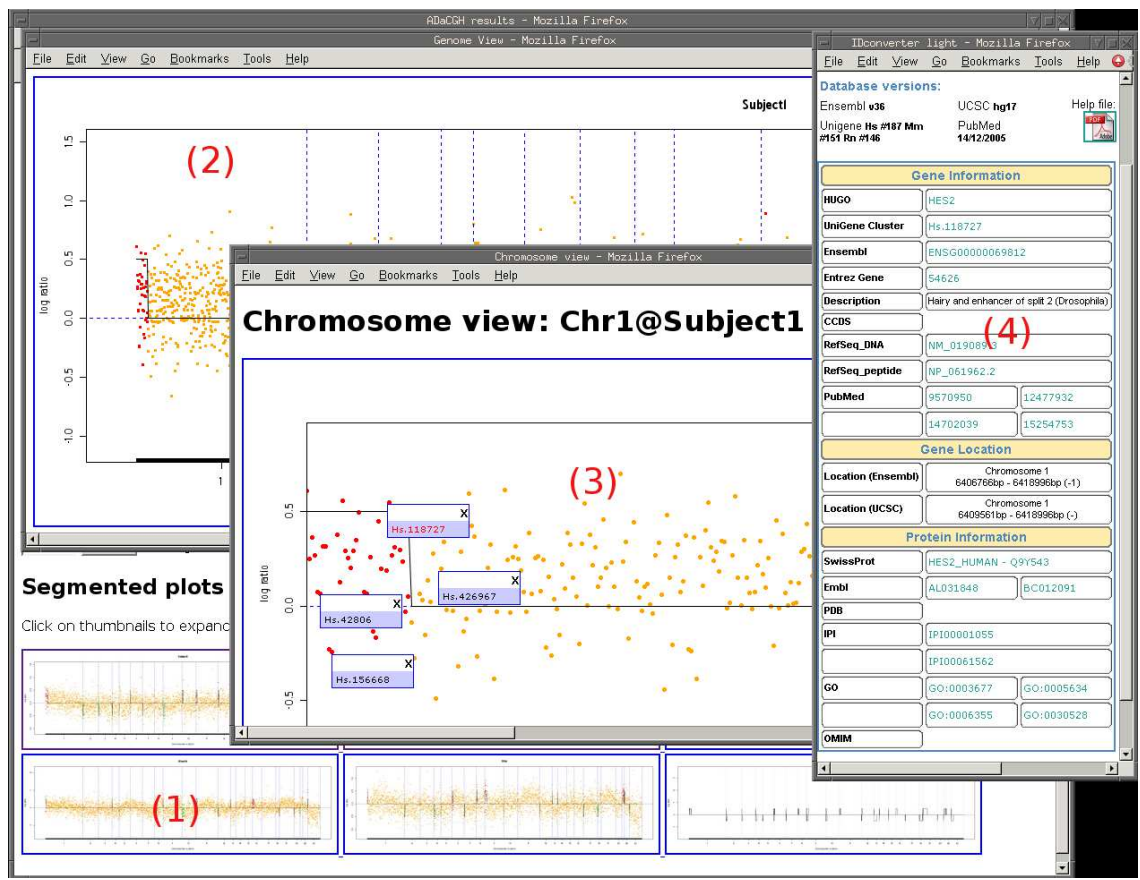}}}

\caption{\label{figure.adacgh} Output from ADaCGH. Partial output from
  ADaCGH showing: (1) the bottom of the main output screen with the
  thumbnails for the segmented plots; (2) Genome View for one of the
  arrays, obtained by clicking on the uppermost thumbnail in (1); (3)
  Chromosome View for the first chromosome (obtained by clicking on the
  region for the first chromosome in (2)), with some data-points showing
  their ID; (4) the results from  IDClight obtained by clicking on the ID
  for one of the highlighted points in (3).}
\end{center}
\end{figure}

\begin{figure}[h!]
\begin{center}

{\resizebox{\columnwidth}{!}{%
\includegraphics{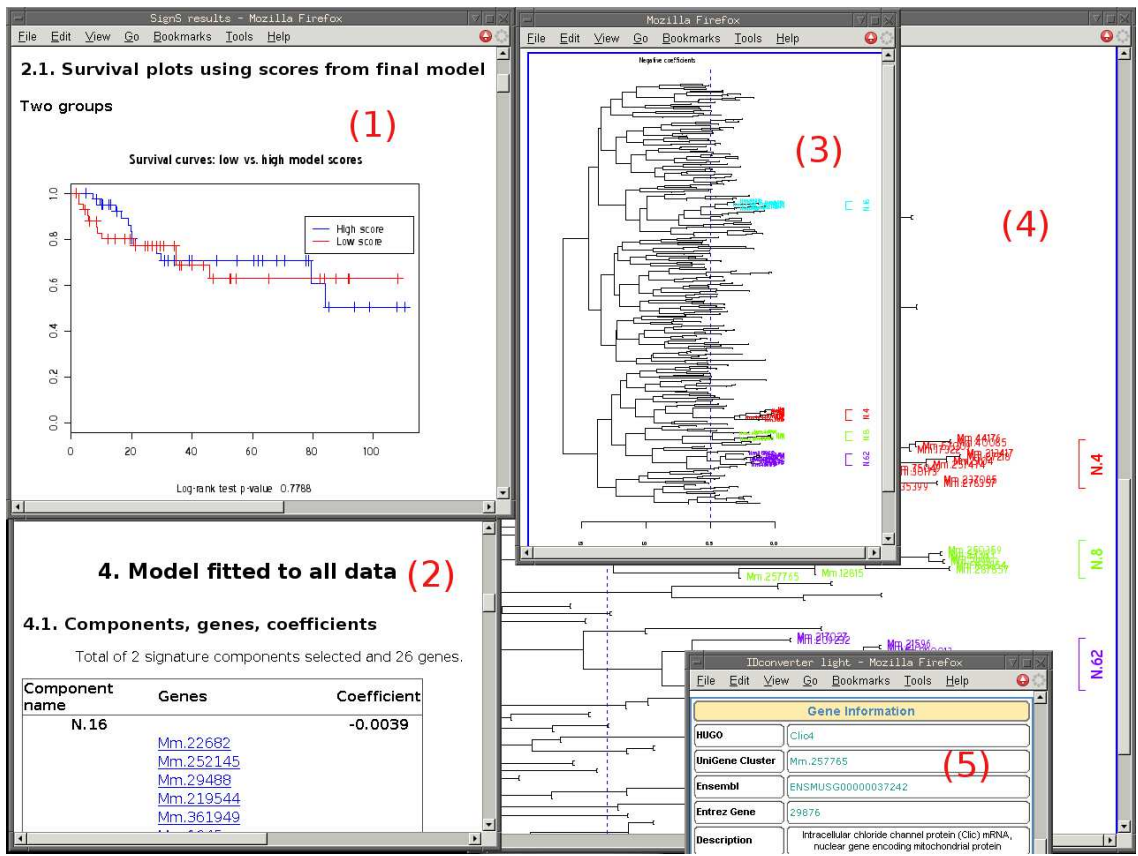}}}

\caption{\label{figure.signs} Output from SignS. Partial output from
  SignS showing: (1) Some model quality plots (survival plots from
  cross-validated prediction scores); (2) part of the output of the model
  fitted to the complete data set; note the clickable gene names; (3) the
  half-size dendrogram for the genes with negative coefficients, showing
  only clusters that fulfill the minimum requirements of correlation and
  size; (4) same as (3), but using the double-size plot; (5) the results
  from clicking, on either (3) or (4), on the cluster leave for Mm.257765.}
\end{center}
\end{figure}

A summary of the output of each application can be seen in Table
\ref{input.output.table}. The complete output from each application, both
tables and figures, can be saved to the user's local file system, thus
allowing for a detailed, complete record of the analysis. As shown in
Figure \ref{asterias.fig}, DNMAD and preP both produce output that can be
sent to other applications of the suite, and SignS, ADaCGH, Pomelo II,
Tnasas and GeneSrF provide clickable tables and figures that call
IDClight.  Examples of output are shown in Figures \ref{figure.adacgh} and
\ref{figure.signs}. 

\subsection{Documentation and help}

All applications have online help, most of them also include tutorials,
detailed and commented examples, and sample data files, and Pomelo II also
has additional tutorials as flash movies.
Tutorials and examples are licensed under a Creative Commons license
(\url{http://www.creativecommons.org}), thus allowing for redistribution
and classroom use. In addition, courses on the use of our tools are taught
occasionally.

\section{Future work}

Our biggest development efforts are currently focused on two areas. We
want to make Asterias easy to deploy at other places; to accomplish this,
we are making available all of the source code as soon as it is ready for
distribution (right now, all the testing code and some application's code
is available from
\url{http://bioinformatics.org/asterias/wiki/Main/DownloadPage} or
\url{https://launchpad.net/projects/asterias}), and we are using a general
purpose web framework (Pylons: \url{http://pylonshq.com}) to ease
distribution and installation. Releasing all of the code and making
installation straightforward might draw other developers into Asterias.
Another current effort focuses on increasing the use of parallelization
and distributed computing to allow for faster responses and more efficient
use of computational resources.

\section{Conclusions}
Asterias is a freely-accessible suite of tools for the analysis of
microarray data, both expression and aCGH, including normalization,
missing data handling and imputation, differential gene expression, class
prediction, survival analysis, and aCGH analysis. Asterias fully exploits
its deployment in a cluster by using web-serving load-balancing and, more
importantly, parallel computing for most of the computationally intensive
tasks.  Asterias also emphasizes sound and tested statistical approaches,
provides careful analysis of the ``multiplicity of solutions'' problem,
and integration of additional functional information.

\section*{Acknowledgments}
We thank the many testers at CNIO and elsewhere that provided feedback on
the applications and bug reports. Funding provided by Fundación de
Investigación Médica Mutua Madrileña and Project TIC2003-09331-C02-02 of
the Spanish Ministry of Education and Science (MEC). R.D.-U. is partially
supported by the Ramón y Cajal programme of the Spanish MEC. Applications
are running on clusters of machines purchased with funds from the RTICCC
from the Spanish FIS.

\raggedright 
\bibliography{asterias}

\bibliographystyle{CancerInformatics} 

\end{document}